\documentclass[twocolumn,floatfix,prc,showpacs,aps,superscriptaddress]{revtex4-1}
\usepackage{graphicx,bm}
\usepackage{amsmath}
\begin{document}

\title{Shell-model half-lives for r-process waiting point nuclei including
first-forbidden contributions
}

\author{Q. Zhi}
\affiliation{Institut f{\"u}r Kernphysik (Theoriezentrum), Technische
  Universit{\"a}t Darmstadt, Schlossgartenstra{\ss}e 2, 64289
  Darmstadt, Germany} 
\affiliation{GSI Helmholtzzentrum f\"ur Schwerioneneforschung,
  Planckstra{\ss}e~1, 64291 Darmstadt, Germany} 
\affiliation{School of Physics and Electronic Science, Guizhou Normal
  University, 550001 Guiyang, PR China}

\author{E. Caurier}
\affiliation{Universit\'e de Strasbourg, IPHC, 23 rue du Loess, 67037
  Strasbourg France CNRS, UMR7178, 67037 Strasbourg, France} 

\author{J. J. Cuenca-Garc\'{i}a}
\affiliation{GSI Helmholtzzentrum f\"ur Schwerioneneforschung,
  Planckstra{\ss}e~1, 64291 Darmstadt, Germany} 

\author{K. Langanke}
\affiliation{GSI Helmholtzzentrum f\"ur Schwerioneneforschung,
  Planckstra{\ss}e~1, 64291 Darmstadt, Germany} 
\affiliation{Institut f{\"u}r Kernphysik (Theoriezentrum), Technische
  Universit{\"a}t Darmstadt, Schlossgartenstra{\ss}e 2, 64289
  Darmstadt, Germany}  
\affiliation{Frankfurt Institute of Advanced Studies, Ruth-Moufang
  Str. 1, 60438 Frankfurt, Germany}

\author{G. Mart\'{\i}nez-Pinedo}
\affiliation{Institut f{\"u}r Kernphysik (Theoriezentrum), Technische
  Universit{\"a}t Darmstadt, Schlossgartenstra{\ss}e 2, 64289
  Darmstadt, Germany}  
\affiliation{GSI Helmholtzzentrum f\"ur Schwerioneneforschung,
  Planckstra{\ss}e~1, 64291 Darmstadt, Germany}
 
\author{K. Sieja}
\affiliation{Universit\'e de Strasbourg, IPHC, 23 rue du Loess, 67037
  Strasbourg France CNRS, UMR7178, 67037 Strasbourg, France} 

\date{\today}

\begin{abstract}
  We have performed large-scale shell-model calculations of the
  half-lives and neutron-branching probabilities of the r-process
  waiting point nuclei at the magic neutron numbers $N=50$, 82, and
  126.  The calculations include contributions from allowed
  Gamow-Teller and first-forbidden transitions. We find good agreement
  with the measured half-lives for the $N=50$ nuclei with charge
  numbers $Z=28$--32 and for the $N=82$ nuclei $^{129}$Ag and
  $^{130}$Cd. The contribution of forbidden transitions reduce the
  half-lives of the $N=126$ waiting point nuclei significantly, while
  they have only a small effect on the half-lives of the $N=50$ and 82
  r-process nuclei.
\end{abstract}
\pacs{21.60.Cs, 23.40.-s, 26.30.Hj}
\maketitle

\section{Introduction}

Although the actual site of the astrophysical r-process is still not
known with certainty, it is commonly accepted that occurs in an
explosive environment of relatively high temperatures ($T \approx
10^9$ K) and very high neutron densities ($> 10^{20}$
cm$^{-3}$)~\cite{BBFH,Cameron,Thielemann91,Thielemann93,woosley}.
Under such conditions, neutron captures are much faster than competing
beta decays and the r-process path in the nuclear chart proceeds
through a chain of extremely neutron rich nuclei with relatively low
and approximately constant neutron separation energies ($S_n \lesssim
3$~MeV). Due to the relatively stronger binding of nuclei with magic
neutron numbers, the neutron separation energies show discontinuities
at the magic numbers $N=50$, 82, and 126.  As a consequence the
r-process matter flow slows down when it reaches these neutron-magic
nuclei and has to wait for several beta decays (which are also longer
than for other nuclei on the r-process path) to occur until further
neutron captures are possible carrying the mass flow to heavier
nuclei. Thus matter is accumulated at these r-process waiting points
associated with the neutron numbers $N=50$, 82, and 126 leading to the
well-known peaks in the observed r-process abundance distribution.

The beta half-lives of the waiting points have at least two
important effects on the r-process dynamics and abundance
distributions. At first, they mainly determine the time it takes the
mass flow within the r-process to transmute seed nuclei to heavy
nuclei in the third peak around $A\sim200$.  Second, in the
astrophysical environment the nuclear r-process timescale (given by
the sum of beta half-lives of nuclei in the r-process path) competes
with some dynamical timescale of the environment, e.g. the expansion
timescale of the ejected matter. If the r-process path and half-lives
were known, the reproduction of the abundance distribution can be used
to constrain the conditions of the astrophysical environment.  If the
r-process has sufficient time for beta-flow equilibrium to establish,
the relative elemental abundances are proportional to the beta
half-life~\cite{Kratz88}.

Despite their importance only a few half-lives of waiting points with
magic neutron numbers $N=50$ and 82 are known
experimentally~\cite{Hosmer,Kratz2,kcd130,Kratz01}, while no
experimental data exist yet for the $N=126$ waiting points. The
situation is expected to improve in the near future with the advent of
new experimental facilities. For example, the beta-decay half-lives of
38 new neutron-rich isotopes from Kr to Tc close to the r-process path
have been measured at the new RIBF facility at
RIKEN~\cite{Nishimura.Li.ea:2011}. Furthermore, researchers at GSI
have measured half-lives of nuclei close to $N=126$ using a novel
analysis method~\cite{Nieto}. Despite this progress, the half-lives
needed for r-process simulations have mainly to rely on theoretical
estimates. As the Q-values involved are rather low, such calculations
have traditionally been based on allowed, i.e. Gamow-Teller,
transitions.  Most of these studies used the Quasiparticle Random
Phase Approximation (QRPA) either on top of semi-empirical global
models \cite{Moeller,Moellerff,ETFSI} or the Hartree-Fock-Bogoliubov
method \cite{Engel}.  Although the calculations give a fair account of
the few experimental half-lives, it is well known that these models
underestimate the correlations among nucleons which pull down the
Gamow-Teller (GT) strength to low energies. This shortcoming is
overcome within the interacting shell model which indeed describes the
measured half-lives of r-process waiting point nuclei very well
\cite{Martinez99,rmp,Cuenca07}.

It is expected that the appearance of intruder single particle states
with different parity may have influence on the low-energy spectra of
the r-process waiting point nuclei. Thus it is conceivable that
first-forbidden transitions might contribute to the half-lives of
these nuclei.  A first attempt to estimate such forbidden
contributions has been taken within the gross
theory~\cite{Moellerff}. This model, however, has been found as rather
inaccurate when applied to Gamow-Teller transitions. More recently,
Borzov extended the QRPA studies based on the Fayans energy functional
to a consistent treatment of allowed and first-forbidden contributions
to r-process half-lives~\cite{Borzov06}. While these calculations find
that forbidden contributions give only a small correction to the
half-lives of the $N=50$ and $N=82$ waiting point nuclei, they result
in a significant reduction of the $N=126$ half-lives.  This important
finding has been our motivation to extend our shell model calculations
of waiting point half-lives to include also first-forbidden
transitions.  We expect that correlations among nucleons will not only
affect the half-lives, but a reliable description of the detailed
allowed and forbidden strength function is needed to estimate the
probabilities for beta-delayed neutron emission rates which are known
to be important to describe the decay of the r-process nuclei towards
stability after freeze-out.

We note that GT and higher multipole transitions are relevant to
describe neutrino-nucleus reactions which are important in many
astrophysical
sites~\cite{Langanke.Martinez-Pinedo:2003,Haxton:2012}. Traditionally,
these reactions have been studied within the Random Phase
Approximation~\cite{Kolbe.Langanke.ea:1992,Kolbe.Langanke.ea:2003},
including nuclei relevant to r-process
nucleosynthesis~\cite{Langanke.Kolbe:2001,Langanke.Kolbe:2002,%
Terasawa.Langanke.ea:2004}. In an interesting recent development,
neutrino-induced reactions on light nuclei with relevance to
neutrino-nucleosynthesis~\cite{Woosley.Hartmann.ea:1990,Heger.Kolbe.ea:2005}
have been calculated on the basis of the shell-model, including GT and
first-forbidden
transitions~\cite{Yoshida.Suzuki.ea:2008,Suzuki.Balantekin.Kajino:2012}.

\section{Shell model and $\beta$  decay theory}

In our half-life calculations, we consider allowed and first-forbidden
contributions. These are obtained using the diagonalization shell mode
code NATHAN developed by Etienne Caurier \cite{Caurier,NATHAN} to
calculate the initial and final nuclear states and the corresponding
nuclear transition matrix elements. Model spaces and residual
interactions will be discussed for the three different sets of waiting
point nuclei with $N=50, 82$ and 126 individually. The partial
half-life, $t$, for a transition between an initial (normally the
ground state) and a final nuclear state is related to the phase space
factor by: 
\begin{equation}
 ft= K = 6146~\mathrm{s}. 
\end{equation}
The phase factor has the form:
\begin{equation}\label{eq:shap}
f=\int_1^{W_0 } {C(W)F(Z,W)(W^2  - 1)^{1/2} } W(W_0  - W)^2 dW\,.
\end{equation}
$C(W)$ is the so called shape factor that depends on the electron
energy, $W$ in units of the electron mass. $W_0$ is the maximum
electron energy, also in electron mass units, that is given by the
difference in nuclear masses between the initial and final nuclear
states, $W_0 = Q/(m_e c^2)= (M_i-M_f)/m_e$. $F(Z,W)$ is the Fermi
function that corrects the phase space integral for the Coulomb
distortion of the electron wave function near the nucleus.  The
partial decay rate is related to the partial half-life: $\lambda = \ln
2/t$. The total decay rate is given by summing over the partial decay
rates to all possible final states.

For allowed transitions, the shape factor does not depend on the
electron energy and for $\beta^-$ decay has the form:
\begin{equation}\label{eq:shap-GT}
C(W) = B(GT).
\end{equation}
The GT reduced transition probability is given by:

\begin{equation}
  \label{eq:bgt}
  B(GT) = \left(\frac{g_A}{g_V}\right)^2
  \frac{\langle f||\sum_k \bm{\sigma}^k \bm{t}^k_- || i
  \rangle^2}{2 J_i +1},
\end{equation}
where the matrix element is reduced with respect to the spin operator
$\bm{\sigma}$ only (Racah convention~\cite{edmons}) and the sum runs
over all nucleons.  For the isospin lowering operator, we use the
convention $\bm{t}_- n = p$. Finally, $(g_A/g_V)=-1.2701(25)$ is the
ratio of weak axial and vector coupling constants.

For first-forbidden (FF) transitions, the shape factor is:
\begin{equation}
  \label{eq:shape-FF}
  C(W) = k + kaW + kb/W + kcW^2
\end{equation}
where the coefficients $k,ka,kb,kc$, depend on the FF nuclear matrix
elements, the maximum electron energy, $W_0$, and the quantity $\xi =
\alpha Z/(2 R)$ with $R$ the radius of a uniformly charged sphere
approximating the nuclear charge
distribution~\cite{Chou.Warburton.Brown:1993}. Following the treatment
of Behrens and B\"uhring~\cite{Behrens} they are given by:

\begin{eqnarray}
  \label{eq:coef}
  k & = & \left[\zeta^2_0+\frac{1}{9}w^2\right]^{(0)}+\left[\zeta_1^2
  +\frac{1}{9}(x+u)^2-\frac{4}{9}\mu_1\gamma_1u(x+u)\right.\nonumber \\
  & & \left. +\frac{1}{18}W_0^2(2x+u)^2 - \frac{1}{18} \lambda_2 (2x-u)^2 \right]^{(1)} \nonumber \\
  & &+ \left[\frac{1}{12}z^2(W_0^2 - \lambda_2)
  \right]^{(2)}, \nonumber\\[0.2cm]
 ka & = & \left[-\frac{4}{3}uY - \frac{1}{9} W_0 (4x^2+5u^2)
 \right]^{(1)} - \left[\frac{1}{6}z^2W_0\right]^{(2)},\\[0.2cm]
 kb & = & \frac{2}{3} \mu_1 \gamma_1 \left\{-\left[ \zeta_0 w
   \right]^{(0)} + \left[\zeta_1 (x+u)\right]^{(1)}\right\},
 \nonumber\\[0.2cm]
 kc & = & \frac{1}{18}\left[ 8u^2 + (2x+u)^2 + \lambda_2 (2x-u)^2\right]^{(1)} \nonumber \\
 & &+ \frac{1}{12} \left[z^2 (1+\lambda_2) \right]^{(2)}.
 \nonumber
\end{eqnarray}
with
\begin{equation}
  \label{eq:coefdef}
  \begin{array}{ll}
    V=\xi^\prime v + \xi w^\prime, & \zeta_0=V+\frac{1}{3} w W_0,\\[2mm]
    Y=\xi^\prime y - \xi (u^\prime+ x^\prime), & \zeta_1= Y +
    \frac{1}{3} (u-x) W_0.
  \end{array}
\end{equation}
The numbers in parenthesis on the closing bracket denote the rank of
the operators inside the bracket. The parameter $\gamma_1$ is given by
$\sqrt{1-(\alpha Z)^2}$. For the Coulomb functions $\mu_1$ and
$\lambda_2$ we use the approximations $\mu_1 \approx 1$ and $\lambda_2
\approx 1$~\cite{Behrensbook}.

After a non-relativistic reduction, the matrix elements can be related
to the form-factor coefficients, $^{A,V}F_{Kls}$, defined in
refs.~\cite{Behrens,Behrensbook}. In the Condon and Shortley
phase convention~\cite{Heydebook} the matrix elements are: 

\begin{subequations}
\begin{eqnarray}
  \label{eq:kabc}
  w & = & -R\ {^AF^0_{011}} \nonumber \\
    & = & -g_A \sqrt{3} \frac{\langle f || \sum_k r_k
  \left[ \bm{C}^k_1 \times \bm{\sigma}^k\right]^0\bm{t}^k_-|| i
  \rangle}{\sqrt{2 J_i +1}} , \\
  x & = & -\frac{1}{\sqrt{3}} R\ {^VF^0_{110}}\nonumber\\
    &  = & - \frac{\langle f||\sum_k r_k
        \bm{C}^k_1 \bm{t}^k_-||i\rangle}{\sqrt{2 J_i +1}} ,\\
  u & = & -\sqrt{\frac{2}{3}}R\ {^AF^0_{111}} \nonumber \\
    & = & -g_A \sqrt{2} \frac{\langle f ||\sum_k r_k \left[ \bm{C}^k_1
      \times \bm{\sigma}^k\right]^1 \bm{t}^k_- || i \rangle}{\sqrt{2 J_i +1}} ,\\
  z & = & \frac{2}{\sqrt{3}} R\ {^AF^0_{211}} \nonumber\\
    & = & 2 g_A \frac{\langle f ||\sum_k r_k \left[ \bm{C}^k_1
        \times \bm{\sigma}\right]^2 \bm{t}^k_- || i \rangle}{\sqrt{2 J_i +1}} ,\\
  w^\prime & = & -\frac{2}{3}R\ {^AF^0_{011}(1,1,1,1)} \nonumber\\
           & = & -g_A \sqrt{3}\frac{\langle f || \sum_k \frac{2}{3} r_k
   I(1,1,1,1,r_k) \left[ \bm{C}^k_1 \times
     \bm{\sigma} \right]^0  \bm{t}^k_- || i \rangle}{\sqrt{2 J_i +1}} ,  \\
  x^\prime & = & -\frac{2}{3\sqrt{3}}R\ {^VF^0_{110}(1,1,1,1)}
               \nonumber\\
           & = & - \frac{\langle f||\sum_k \frac{2}{3} r_k I(1,1,1,1,r_k)
           \bm{C}^k_1 \bm{t}^k_- ||i \rangle}{\sqrt{2 J_i +1}} ,\\
  u^\prime & = & -\frac{2\sqrt{2}}{3\sqrt{3}}R\
           {^AF^0_{111}(1,1,1,1)}\nonumber\\ & = & -g_A \sqrt{2}
           \frac{\langle f||\sum_k \frac{2}{3}  r_k I(1,1,1,1,r_k) \left[
             \bm{C}^k_1 \times \bm{\sigma}^k \right]^1
           \bm{t}^k_- || i \rangle}{\sqrt{2 J_i +1}} ,\\
           \xi^\prime v & =  &  {^AF^0_{000}} \nonumber \\
            & = & \frac{g_A \sqrt{3}}{M} \frac{\langle f ||\sum_k
            \left[\bm{\sigma}^k \times \bm{\nabla}^k\right]^0 \bm{t}^k_-
            || i \rangle}{\sqrt{2 J_i +1}},\\
            \xi^\prime y & = & {^VF^0_{101}} \nonumber \\
            & =& -\frac{1}{M} \frac{\langle f ||\sum_k
            \bm{\nabla}^k \bm{t}^k_-
            || i \rangle}{\sqrt{2 J_i +1}},
         \end{eqnarray}
       \end{subequations}
where 
\begin{equation}
  \label{eq:Cl}
  \bm{C}_{lm}=\sqrt{\frac{4\pi}{2l+1}} \bm{Y}_{lm},
\end{equation}
with $\bm{Y}_{lm}$ the spherical harmonics. The weak axial coupling
constant is $g_{A}=-1.2701(25)$ and $M$ is the nucleon mass. The
quantity $I(1,1,1,1,r)$ appearing in the primed matrix elements takes
in account the nuclear charge distribution that can be approximated by
a uniform spherical distribution~\cite{Behrens}:

\begin{equation}
  \label{eq:Ir}
  I(1,1,1,1,r)= \frac{3}{2}\left\{
    \begin{array}{cc}
      \displaystyle{\left[1-\frac{1}{5}\left(\frac{r}{R}\right)^2\right]} &
      0\leq r \leq R,\\[0.5cm]
      \displaystyle{\left[\frac{R}{r}-\frac{1}{5}\left( \frac{R}{r}
          \right)^3 \right]} & r\geq R.
    \end{array}\right.
\end{equation}

Based on the conserved vector current theory and the assumption that
the isospin is a good quantum number, the 
matrix element $\xi'y$ can be related to the $x$ matrix
element~\cite{Warburtonap}:  
\begin{equation}
 \xi'y=E_{\gamma}x,
\end{equation}
where energy $E_{\gamma}$ is defined as the energy difference between
the isobaric analog of initial state and the final state: 

\begin{equation}
  \label{eq:2}
  E_\gamma = E_{\text{ias($i$)}} - E_f = Q + \Delta E_C -(m_n c^2
  - m_p c^2), 
\end{equation}
where $m_n$ and $m_p$ are the neutron and proton masses and $\Delta
E_C$ is the Coulomb displacement energy between isobaric analog states
that can be approximated by~\cite{Antony.Pape.Britz:1997}:

\begin{equation}
  \label{eq:ecoul}
  \Delta E_C = 1.4136(1)\ \bar{Z}/A^{1/3} - 0.91338(11)~\text{MeV},
\end{equation}
with $\bar{Z}=(Z_i+Z_f)/2$.

To compare the first-forbidden and 
Gamow-Teller transitions, we define  the averaged shape factor:
\begin{equation}
  \label{eq:C4}
\overline {C(W)}= f/f_0,
\end{equation}
where $f$ takes the form of Eq. \ref{eq:shap} and $f_0$ is
\begin{equation}
f_0=\int_1^{W_0 } {F(Z,W)(W^2  - 1)^{1/2} } W(W_0  - W)^2 dW.
\end{equation}


\section{Model spaces and quenching}

We have performed beta-decay half-live calculations for r-process
waiting points based on large-scale shell model calculations. In
particular we chose the following model spaces and respective
interactions.

For the $N=50$ nuclei we have
adopted a model space spanned by the $0f_{7/2,5/2}$ and $1p_{3/2,1/2}$
orbits for protons and by the $0f_{5/2}$, $1p_{3/2,1/2}$, and $0g_{9/2}$
orbits for neutrons. The single-particle energies and the
residual interaction are the ones adopted in  \cite{Sieja10}
to study the shell evolution between $^{68}$Ni and $^{78}$Ni.

Our shell model calculations for the $N=82$ waiting point nuclei
follows the shell model studies presented in \cite{Cuenca07}.  From
the two model spaces defined in \cite{Cuenca07} we adopt the one built
on a $^{88}$Sr core. That is we explicitly consider the $1p_{1/2}$
proton orbit which is expected to be important for the description of
the negative parity states and hence the first-forbidden
transitions. Our model space is then spanned by the
$0g_{7/2},1d_{3/2,5/2},2s_{1/2},0h_{11/2}$ orbits outside the $N=50$
core for neutrons, and the
$1p_{1/2},0g_{9/2,7/2},1d_{3/2,5/2},2s_{1/2}$ orbits for protons.
This model space avoids spurious center-of-mass excitations by
omitting the $0h_{11/2}$ orbit for protons and the $0g_{9/2}$ orbit for
neutrons. We adopt the residual interaction given in \cite{Cuenca07}
based on $^{88}$Sr core which gives a good account of the spectroscopy
of nuclei in the neighborhood of $^{132}$Sn. In particular, our
calculation reproduces the excitation energy of the first $1^+$ state
in $^{130}$In as well as the low-energy spectrum of $^{128}$Cd and of
the r-process waiting point nucleus $^{130}$Cd \cite{Jungclaus}.

The model space for the $N=126$ waiting points has been spanned by the
$0g_{7/2}$, $1d_{5/2,3/2}$, $0h_{11/2}$ and $2s_{1/2}$ orbits for
protons and the $0h_{9/2}$, $1f_{7/2,5/2}$, $0i_{13/2}$,
$2p_{3/2,1/2}$ orbits for neutrons. As interaction we use the
effective Kuo-Herling interaction KHH$_e$ of ref.~\cite{Warburton91}
which has been constructed based on holes in a $^{208}$Pb core. It is
the same model space and effective interaction as has been used in a
previous calculation of the half-lives, which, however, has only
considered pure GT transitions \cite{Langanke.Martinez-Pinedo:2003}.
These are mainly connected to neutron $0h_{9/2}$ to proton $0h_{11/2}$
transitions. However, it is expected that first-forbidden transitions
can compete, mainly via neutron $0i_{13/2}$ to proton $0h_{11/2}$
transitions. Full diagonalization in this model space exceeds current
computer capabilities. Hence, we performed truncated calculations
following a generalize seniority scheme that allows for configurations
with maximum seniority 8, i.e. we consider a maximum of 4 non-$J=0$
pairs, for even even nuclei.  For odd-even nuclei, the number of
broken pairs had to be limited to three, while for $^{199}$Ta no
limitation has been enforced.  We expect our model spaces to be large
enough to give a reasonable account for the low-lying Gamow-Teller and
first-forbidden transitions. Nevertheless the model spaces are too
restricted to recover the full Gamow-Teller and first-forbidden
strengths built on the ground or isomeric states.  The missed
strength, however, resides mainly outside the $Q_\beta$ window and
does hence not affect our half-life calculations.

Although the shell model usually gives a good account of the relative
strength distributions, it overestimates the total strength. For
Gamow-Teller transitions this shortcoming can be corrected for by
replacing the bare Gamow-Teller operator by an effective operator
GT$_{\rm eff}$ = $q$ $\bm{\sigma t}$. The quenching factor $q$ has been
found to be approximately constant over the nuclear chart
\cite{Wildenthal,Langanke95,Martinez96}. In practice, using $q \approx
0.7$ has been shown to give a good reproduction of the absolute
Gamow-Teller distributions. There is evidence that also the absolute
first-forbidden transition strength is overestimated within shell
model approaches. Ejiri and collaborators \cite{Ejiri72,Ejiri78}
related this fact to core polarization effects and suggested the
introduction of a constant hindrance factor.  Based on perturbation
theory, Warburton \cite{Warburton,Warburton91} showed that the
quenching of first-forbidden transitions depends slightly on the
initial and final single particle orbits.  In particular, Warburton
found that transitions mediated by the rank 0 operator (Eq. 8h) (called
relativistic matrix element) appear to be enhanced compared to the
other first-forbidden transitions, due to meson exchange
effects~\cite{Warburton94}. 
 
\begin{figure}[htb]
  \includegraphics[width=\linewidth]{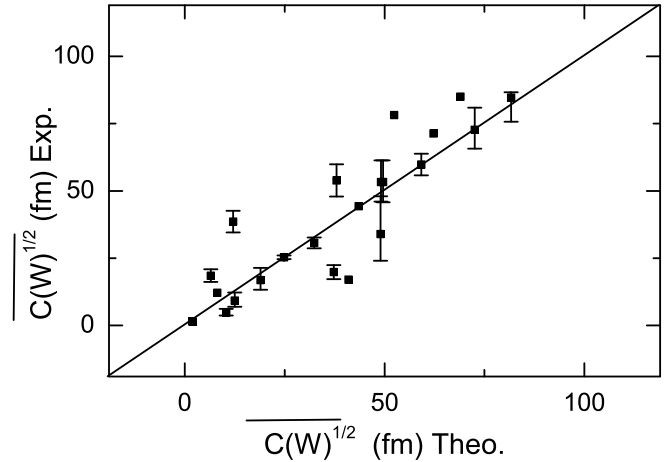}
  \caption{Comparison of calculated first-forbidden average shape
    factors, obtained for the best fit values of the quenching factors
    (Eq. (16)), with experimental data \cite{Fog04,Geer80,Nndc}.}
    \label{fig:shape}
\end{figure}

To treat the quenching of the Gamow-Teller and first-forbidden
transitions in our shell model calculations we have assumed that the
quenching factors are the same for all nuclei. Following the findings
of Warburton \cite{Warburton,Warburton94} we have furthermore assumed
that the quenching factors for the operators of rank 0, 1, and 2
contributing to the first-forbidden transitions as defined in Eq. (8)
can be different. To determine these individual quenching factors we
have performed shell model calculations for experimentally known
beta-decays of nuclei in the vicinity of the magic neutron numbers
$N=82$ and $N=126$.  Here, we have adopted the similar set of
first-forbidden transitions of nuclei in the lead region as chosen in
the study of Warburton \cite{Warburton94}, supplemented by the decays
of the ground state of $^{205}$Au ($N=126$) and the $(1/2)^-$ isomeric
states in $^{131}$In ($N=82$) and $^{129}$In ($N=80$) which are both
known to decay by first-forbidden transitions. By performing a
least-squares fit to the experimental data we obtained the following
quenching factors for the various matrix elements defined in Eq. (8):
\begin{equation}
  \label{eq:1}
  \begin{array}{ll}
    q(\xi^\prime v)=1.266, & q(w)=q(w^\prime)=0.66, \\
    q(x)=q(x^\prime) = 0.51, & q(u)=q(u^\prime)=0.38, \\
    q(z)=0.42. \\
  \end{array}
\end{equation}
The calculated half-lives and the corresponding average shape factors
are summarized in Table~\ref{tab:shape}. Fig. \ref{fig:shape} compares
the experimental and calculated shape factors.

\begin{table}[bth]
  \caption{Comparison of calculated $\log f_0 t$ and
    $\left(\overline{C(W)}\right)^{1/2}$ for first-forbidden  transitions with
    experimental data~\cite{Fog04,Geer80,Nndc}.}
   \label{tab:shape}
   \renewcommand{\arraystretch}{1.1}
  \begin{ruledtabular}
  \begin{tabular}{cccccc}
\multicolumn{2}{c}{Transition} & \multicolumn{2}{c}{log $f_0 t$}
&\multicolumn{2}{c}{$\left(\overline {C(W)}\right)^{1/2}$} \\ \cline{1-2}\cline{3-4}\cline{5-6}
 Initial &Final  &the  & exp &the  & exp \\ \hline
$^{131}$In($\frac{1}{2}^-$)   &$^{131}$Sn($\frac{3}{2}^+$)     &5.32 &$\approx$5.1 &65.8 &85.4  \\
                              &$^{131}$Sn($\frac{1}{2}^+$)     &5.74 &6.5          &41.1 &17.1 \\
$^{129}$In($\frac{1}{2}^-$)   & $^{129}$Sn($\frac{3}{2}^+$)    &5.57 &5.9(3)       &49.8 &34(12)\\
                              & $^{129}$Sn($\frac{1}{2}^+$)    &5.80 &5.5(1)       &38.1 &54(6) \\
$^{205}$Hg($\frac{1}{2}^-$)   & $^{205}$Tl($\frac{1}{2}^+_1$)  &5.37 &5.257(11)    &62.3 &71.3(9)\\
                              & $^{205}$Tl($\frac{1}{2}^+_2$)  &6.77 &7.03(25)     &12.5 &9(3)\\
                              & $^{205}$Tl($\frac{3}{2}^+_1$)  &7.26 &6.51(21)     &18.9 &17(4)\\
                              & $^{205}$Tl($\frac{3}{2}^+_2$)  &6.32 &7.61(22)     &10.3 &5(1)\\
                              & $^{205}$Tl($\frac{5}{2}^+$)    &8.16 &8.70(21)     &1.91 &1.3(3)\\
$^{206}$Hg($0^+$)             & $^{206}$Tl($0^ -$)             &5.42   &5.41(6)    &59.2 &60(4) \\
                              & $^{206}$Tl($1_1^-$)            &5.18   &5.24(10)   &77.6 &73(8) \\
                              & $^{206}$Tl($1_2^ -$)           &5.68   &5.67       &43.6 &44.3\\
$^{207}$Tl($\frac{1}{2}^+$)   & $^{207}$Pb($\frac{1}{2}^-$)    &5.14   &5.108(6)   &81.7 &84.5(6)\\
                              & $^{207}$Pb($\frac{3}{2}^-$)    &6.18   &6.157(22)  &24.7 &25.3(6)\\
$^{206}$Tl($0^-$)             & $^{206}$Pb($0_1^ +$)           &5.42   &5.1775(13) &52.4 &78.0(1)\\
                              & $^{206}$Pb($0_2^ +$)           &5.18   &5.99(6)    &32.4 &31(2) \\
                              & $^{206}$Pb($2^ +$)             &5.68   &8.60(3)    &1.87 &1.52(5)\\
$^{205}$Au($\frac{3}{2}^+$)   & $^{205}$Hg($\frac{1}{2}^-_1$)  &6.79   &5.79(9)    &12.1 &39(4)\\
                              &$^{205}$Hg($\frac{3}{2}^-_1$)   &7.33   &6.43(11)   &6.5  &18(2)   \\
                              &$^{205}$Hg($\frac{5}{2}^-_1$)   &5.82   &6.37(12)   &37.3 &20(3)   \\
   \end{tabular}
  \end{ruledtabular}
\end{table}

Upon a closer inspection there is quite a good agreement between our
shell model half-lives for the isomeric states in the In isotopes with
the experimental values.  For the nuclei in the vicinity of $N=126$ we
find, however, a noticeably larger scatter between calculation and
data. Satisfyingly there are no systematic deviations.  With the
exception of the two $^{205}$Au decays, where our calculation
overestimates (to the $5/2^-$ state in $^{205}$Hg) or underestimates
(to the $1/2^-$ state) the average shape factor roughly a factor 9, we
generally find agreement of our calculated $\bar{C(W)}$ with data
within a factor of 4.  As already observed while determining the
quenching factor for GT transitions in shell model calculations
\cite{Martinez95} the description of a decay between specific states
is noticeably more sensitive to nuclear structure effects than global
quantities like half-lives or total strengths. Hence we expect that
our prescription of quenching for first-forbidden transitions yields a
fair description of the $N=126$ half-lives.
  
As already stressed by Warburton \cite{Warburton94} the relativistic 
matrix element (Eq. 8h) is enhanced compared to the other first-forbidden
transitions. We confirm this finding as the value for $q(\xi^\prime v)$ is
noticeably larger than the other quenching factors.

Having determined the quenching of first-forbidden transitions, we adjust
the quenching of the Gamow-Teller transition to the half-life of 
$^{130}$Cd which is expected to decay dominantly by Gamow-Teller.
This is indeed borne out in our calculation.
Using the quenching factor $q_{GT}=0.66$ we reproduce the measured
half-life using both GT and first-forbidden transitions. The latter contribute
about $13\%$ to the half-life and hence are small, but not negligible.
Our factor $q_{GT}$ is only slightly smaller than the customary
quenching value of 0.7.

All half-lives presented in the next sections for the r-process
waiting point nuclei have been obtained using the quenching factors
for Gamow-Teller and first-forbidden transitions derived above.

Before we present our results we have to discuss another potential
shortcoming of our calculations of first-forbidden transitions and how
we will handle it.  A completely converged calculation of the
first-forbidden transition strength in our chosen model space is
prohibited due to computational limitations. We have derived the
strength within the Lanczos scheme using 100 iterations. As a
consequence the lowest states are converged and correspond to physical
states, while the Lanczos states at higher excitation energies are
unphysical and represent strength per energy interval.  Furthermore,
their energy positions depend on the sum rule (pivot) state used for
the calculation of the strength function. For example, starting from
the pivot state of the $x$ operator one obtains different energy
positions for the non-converged states than starting from the sum rule
state of the $u$ operator.  As we need to compute superpositions of
operators like $u+x$, where both the magnitude and the phase of the
individual operators matter, we have followed the same procedure as is
used in shell-model calculations of double-beta decays
\cite{Caurier}. Hence we start with an arbitrary sum rule state that
can be any linear combination of operators of the same rank. During
the Lanczos iteration procedure we compute the overlaps with the
individual operators. This iteration procedure is stopped when at
least $80\%$ of the total strength for each individual operator is
recovered.


To illustrate this point, we have performed calculations for
$^{199}$Ta in the model space defined above and have used two
different linear combinations of rank 1 operators, $\xi ^\prime y -
\xi (x + u)$ and $(x + u)$, as pivots for the Lanczos calculations of
the operators $x$ and $u$. For the combination $\xi ^\prime y - \xi (x
+ u)$ we obtain $92\%$ and $80\%$ of the total strength for the
operators $x$ and $u$, respectively, while for the combination $x+u$
we recover $85\%$ and $95\%$ after 100 iterations.  We stress that the
effect which these shortcomings have on the first-forbidden half-lives
is mildened as the contributions of the low-lying states, which are
converged in our Lanczos scheme, are strongly enhanced by the phase
space energy dependence.

\begin{figure}[htb]
    \includegraphics[width=\linewidth]{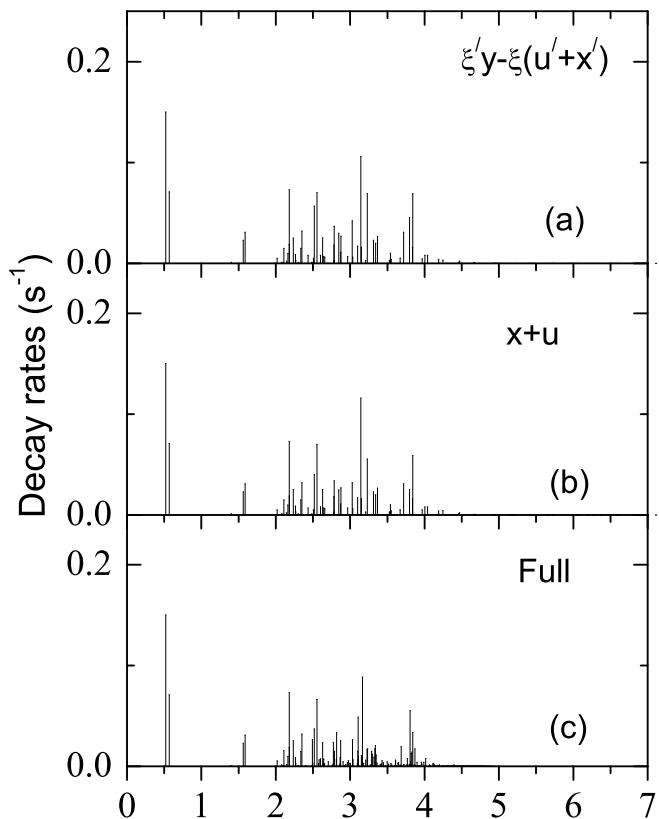}
    \caption{Partial decay rates for $^{199}$Ta calculated from
      first-forbidden rank 1 transitions only. The top (a) and
      medium (b) panel shows the decay rates obtained by using the
      linear combinations $\xi ^\prime y - \xi (x + u)$ and $(x + u)$,
      respectively, as pivot elements to calculate the contributions
      of rank 1 operators within a Lanczos scheme with 100 iterations.
      The bottom panel (c) shows the partial decay rate using a Lanczos
      scheme with 300 iterations in which the states within the
      $Q_\beta$ window are converged. These shell model calculations
      have been performed in a truncated model space compared to other
      studies of the $N=126$ isotones.}
    \label{fig:converged-Ta}
\end{figure}

To quantify the potential uncertainty in our first-forbidden
half-lives, we have performed again calculations for $^{199}$Ta in the
model space as defined above, however, allowing only one proton pair
to be broken in our seniority scheme. This truncated space allows for
the calculation of a fully converged first-forbidden strength
distribution in the $Q_\beta$ window. Fig.  \ref{fig:converged-Ta}
compares the partial decay rates to the various states in the daughter
nucleus obtained in the fully converged calculation (with 300 Lanczos
iterations) with those where the rank 1 contributions were derived
using the linear combinations $\xi ^\prime y - \xi (x + u)$ and $(x +
u)$ as pivots for the Lanczos scheme with 100 iterations. As expected,
the lowest Lanczos states are converged and hence the calculated
strength is the same for either choice of pivot combinations. The
Lanczos scheme with 100 iterations is not sufficient to converge the
states for excitation energies larger than about 2.5 MeV. They
represent unphysical states as discussed above.  Obviously further
iterations lead to a stronger fragmentation of the strength in a small
energy interval around the unphysical states. The energy interval is
small enough that this redistribution of strength due to different
phase space weighting has negligible effect on the half-life. But also
the interference contributions lead to rather mild differences between
the truncated and the converged calculations. We find a partial
half-life due to rank 1 first-forbidden operators of 655~ms and 716~ms
when calculating the rank 1 operators from the linear combinations
$\xi ^\prime y - \xi (x + u)$ and $(x + u)$, respectively, while the
partial half-life in the converged study is 651 ms agreeing within 10 percent
with the two approximate calculations.  In the following we will
calculate the contributions from the first-forbidden rank 1 operators
using a Lanczos scheme with the pivot state $\xi ^\prime y - \xi (x +
u)$ and 100 iterations.

\section{Half-lives of the $N=50$ waiting point nuclei}

\begin{figure}[htb]
    \includegraphics[width=\linewidth]{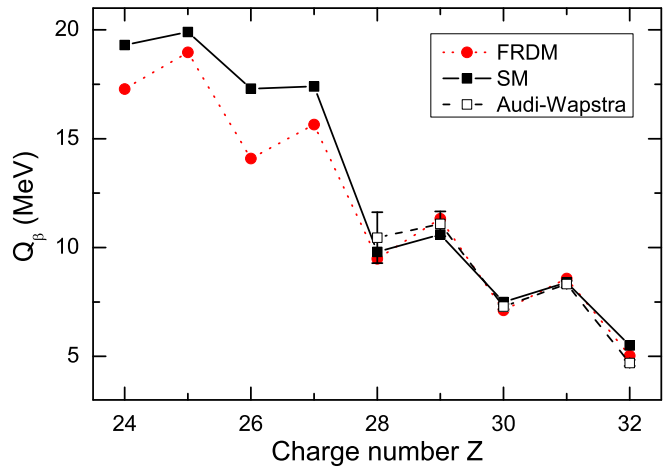}
    \caption{(Color online) Comparison of $Q_{\beta}$ values of the
      $N=50$ isotones between experimental data \cite{Audi03} and
      theoretical results from FRDM~\cite{Moellerff} and the present
      shell model approach.}
    \label{fig:Q50}
\end{figure}

To calculate half-lives, a good description of the transition matrix
elements and also of the $Q_\beta$ values is required.  As is
demonstrated in Fig. \ref{fig:Q50} our shell model calculation
reproduces the $Q_\beta$-values as given in the Audi-Wapstra
compilation well \cite{Audi03}. Hence we will use the shell model
$Q_\beta$ values in the following calculation of the half-lives and
$\beta$-delayed neutron emission probabilities for the $N=50$ waiting
point nuclei.

\begin{table}[htb]
  \caption{Comparison of the present shell model
    half-lives and the ones of
    reference~\cite{Langanke.Martinez-Pinedo:2003}. 
    All half-lives are in s.}
  \label{tab:lives50}
  \renewcommand{\arraystretch}{1.1}
  \begin{ruledtabular}
    \begin{tabular}{cccc}
      Nucleus & \multicolumn{3}{c}{Half-Life (s)} \\ \cline{2-4}
      &   Expt.  & \textrm{present} & \textrm{shell model
        (ref~\cite{Langanke.Martinez-Pinedo:2003})} \\ \hline
      $^{82}$Ge  & $4.55 \pm 0.05$        & 6.90                  & 2.057                  \\
      $^{81}$Ga  & $1.217 \pm 0.005$      & 1.03                  & 0.577                  \\
      $^{80}$Zn  & $0.545 \pm 0.016$      & 0.53                  & 0.432                  \\
      $^{79}$Cu  & $0.188 \pm 0.025$      & 0.27                  & 0.222                  \\
      $^{78}$Ni  & $0.11^{+0.1}_{-0.06}$  & 0.15                  & 0.127          
      \\
      $^{77}$Co  &                        & 0.016                 & 0.016                  \\
      $^{76}$Fe  &                        & $8.26 \times 10^{-3}$  & $7.82\times 10^{-3}$    \\
      $^{75}$Mn  &                        & $3.66 \times 10^{-3}$  & $3.52\times 10^{-3}$    \\
      $^{74}$Cr  &                        & $2.23 \times 10^{-3}$  & $2.07\times 10^{-3}$    \\
    \end{tabular}
  \end{ruledtabular}
\end{table}

\begin{figure}[htb]
    \includegraphics[width=\linewidth]{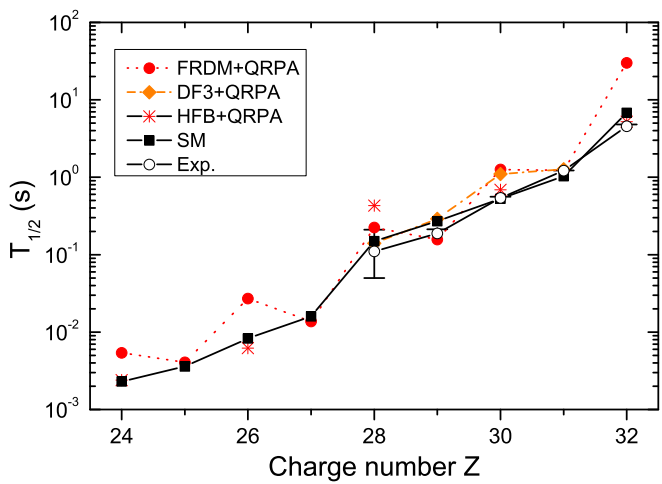}
    \caption{(Color online) Comparison of half-lives of the $N=50$
      isotones between experimental data and theoretical results from
      FRDM+QRPA~\cite{Moellerff}, HFB+QRPA \cite{Engel}, DF3+QRPA
      \cite{Borzov,Borzov06} and the present shell model approach.}
    \label{fig:lives50}
\end{figure}

As is shown in Table \ref{tab:lives50} and in Fig. \ref{fig:lives50}
the shell model half-lives agree quite well with the data, although
they overestimate the ones of $^{82}$Ge and $^{79}$Cu by about $50 \%$.
Nevertheless the agreement is significantly better than obtained based
on the global FRDM and ETFSI models. The HFB results \cite{Engel},
which are restricted to the decay of even-even nuclei,
are very similar to the shell model results, except for the
half-life of the double-magic nucleus $^{78}$Ni. Here only the shell model
reproduces the measured value \cite{Hosmer}, while all other models
predict a significantly longer half-life. This underlines again the fact
that many-body configuration mixing is needed to reproduce the cross-gap
correlations in double-magic nuclei. Similar results have been found
in studies of the isotope shifts in calcium~\cite{caurier01} or the
M1 strength distributions in argon isotopes~\cite{Lisetskiy}.

\begin{figure}[htb]
    \includegraphics[width=\linewidth]{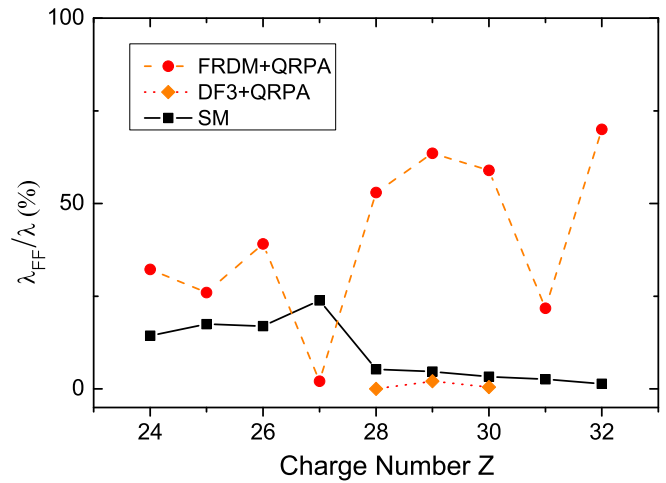}
    \caption{(Color online) Percentage of the contributions from
      first-forbidden transitions to the half-lives of the N$=50$
      isotones from FRDM+QRPA \cite{Moellerff}, DF3+QRPA
      \cite{Borzov,Borzov06} and the present shell model.}
    \label{fig:prob50}
\end{figure}

The contribution of first-forbidden transitions to the $N=50$
half-lives is shown in Fig. \ref{fig:prob50}.  For the decay of the
nuclei with $Z \geq 28$ the probability is very small (less than $5
\%$). However, first-forbidden transitions contribute about $25 \%$ to
the $^{77}$Co decay, while they are smaller, but still sizable for the
decay of the nuclei with charge numbers $Z=24$--26.

\begin{figure*}[tb]
    \includegraphics[width=\linewidth]{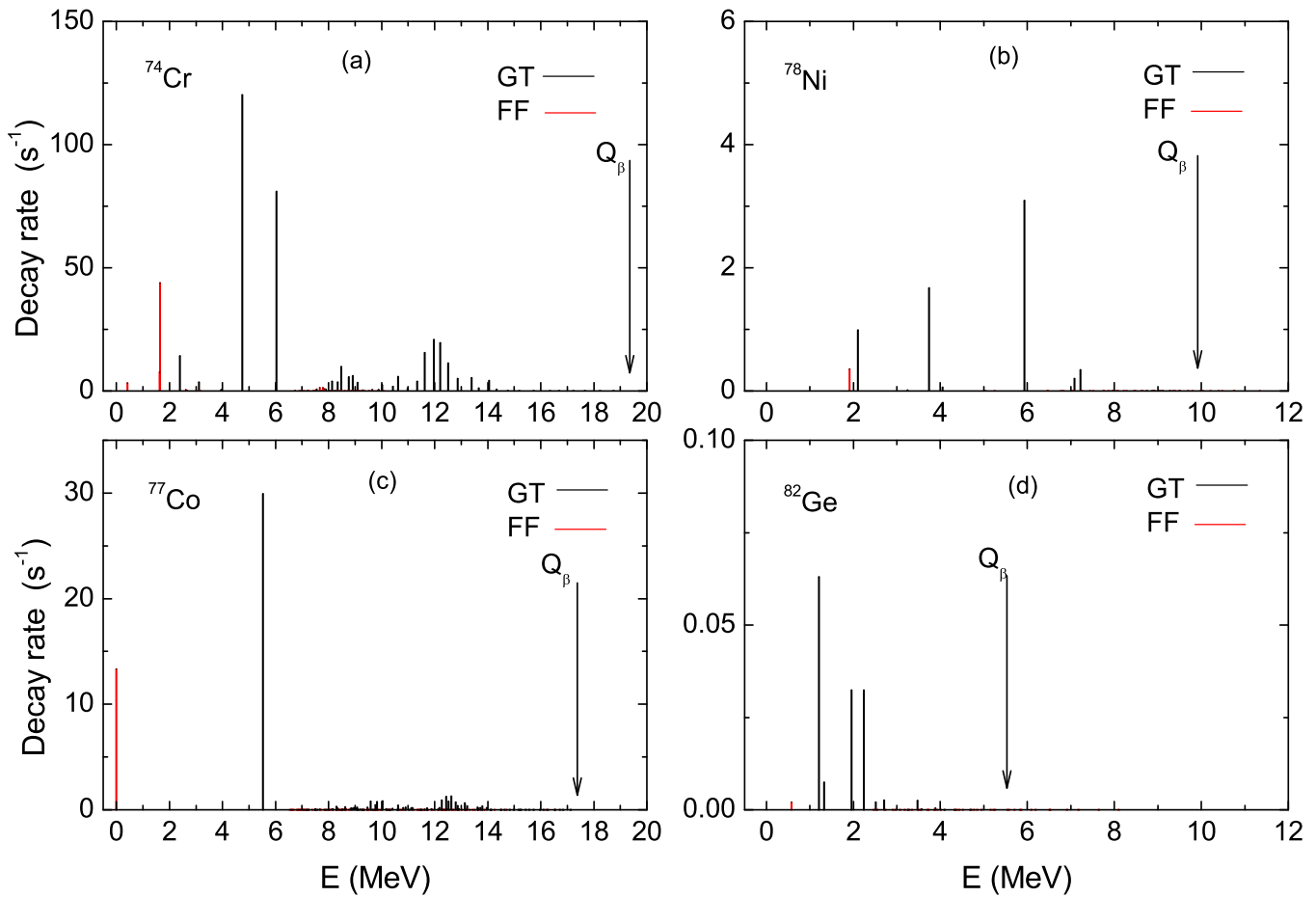}
    \caption{(Color online) Partial decay rates including GT and FF
      transitions for the $N=50$ isotones $^{74}$Cr (a), $^{78}$Ni
      (b), $^{77}$Co (c), and $^{82}$Ge (d).}
    \label{fig:rates-50}
\end{figure*}

To understand this behavior, we note that first-forbidden
contributions are related to the transition from a $g_{9/2}$ neutron
orbital to a $f_{7/2}$ proton orbital for the rank 1 operators and to
a $f_{5/2}$ proton orbital for rank 2 operators.  (There are no
contributions from rank 0 operators in our model space.)  In the
simple Independent Particle Model the $f_{7/2}$ level gets completely
occupied for $^{78}$Ni and consequently this transition is
Pauli-blocked for $N=50$ nuclei with $Z \ge 28$. In the shell model,
the blocking is partially removed by configuration mixing, but the
importance of the first-forbidden transitions stay low.  For these
nuclei they are nearly exclusively due to contributions from the rank
2 operators.  For the nuclei with $Z <28$ the proton $f_{7/2}$ orbital
is not fully occupied and first-forbidden transitions due to the rank
1 operators are possible. They are noticeably larger those of the rank
2 operators. Hence the total first-forbidden strength is significantly
larger for nuclei with $Z < 28$ than for the nuclei with $Z \ge
28$. Furthermore it increases with decreasing charge number due to the
depopulation of the proton $f_{7/2}$ orbital in the daughter
nucleus. However, also Gamow-Teller transitions from the neutron
$f_{5/2}$ orbital into the proton $f_{7/2}$ orbital become unblocked.
Hence reducing the charge number, increases both the GT and
first-forbidden transitions due to decreasing Pauli-blocking of the
dominant transitions into the $f_{7/2}$ orbital.  However, the
relative decrease for the GT half-lives with decreasing charge number
is stronger than for the first-forbidden transitions.  This is related
to phase space. Examples of differential decay rates as function of
excitation energies for nuclei $^{74}$Cr, $^{77}$Co, $^{78}$Ni and
$^{82}$Ge are shown in Figs.~\ref{fig:rates-50}. From these figures we
note that the first-forbidden transitions are dominantly proceeding to
states in the daughter at low excitation energies (usually up to 2-2.5
MeV) for the $N=50$ nuclei with $Z<28$, while the GT transitions go to
states with excitation energies of order 5-7 MeV, simply reflecting
the fact that it is energetically more favorable to have a $f_{5/2}$
neutron hole and a closed $g_{9/2}$ shell, than having a hole in the
$g_{9/2}$ orbital.  As the energy gain in the transitions is smaller
for the GT transitions, they are more sensitive to the increase of the
$Q_\beta$ value with decreasing charge number. This explains why the
relative contribution of first-forbidden transitions decreases with
reduced charge number below the double-magic $^{78}$Ni.  Above
$^{78}$Ni the GT transitions proceed to daughter states at relatively
low excitation energies. (Fig. \ref{fig:rates-50} shows the
differential decay rates for $^{82}$Ge as an example.) As a
consequence first forbidden transitions, due to their smaller
transition matrix elements, cannot compete with GT transitions.

\begin{figure}[htb]
    \includegraphics[width=\linewidth]{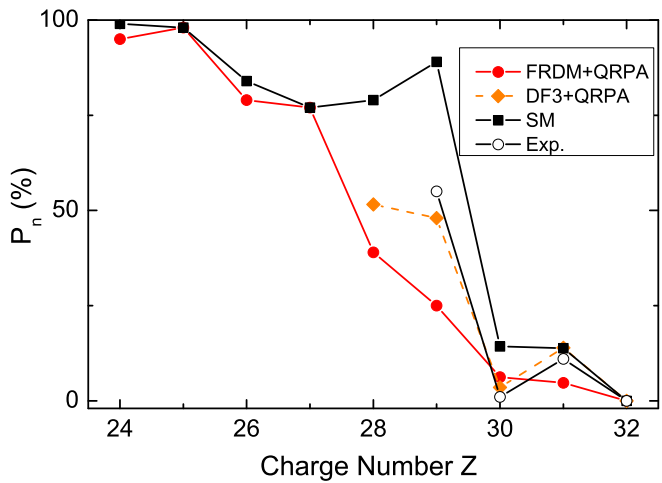}
    \caption{(Color online) Neutron emission probability for N$=50$
      isotones from FRDM+QRPA \cite{Moellerff}, DF3+QRPA
      \cite{Borzov,Borzov06}, the present shell model and experiment
      \cite{Nndc}.}
    \label{fig:neut50}
\end{figure}

We have used the shell model neutron separation energies to calculate
the $\beta$-delayed neutron emission probabilities, which are shown in
Fig. \ref{fig:neut50}. As demonstrated above GT transitions dominate
the decays of the $N=50$ r-process waiting points. As for the nuclei
with $Z \leq 29$ these transitions connect mainly to daughter states
at excitation energies above the respective neutron threshold, the
decay is accompanied by neutron emission with a high probability. The
probability comes close to $100 \%$ for the decays of $^{55}$Mn and
$^{54}$Cr associated with the very low neutron separation energies in
the daughters. For the nuclei with $Z>29$, the GT transitions reside
at lower excitation energies in the respective daughters, while the
neutron thresholds increases. As a result the neutron emission
probability is strongly reduced in these nuclei.

\section{Half-lives of the $N=82$ waiting point nuclei}

\begin{figure}[htb]
  \includegraphics[width=\linewidth]{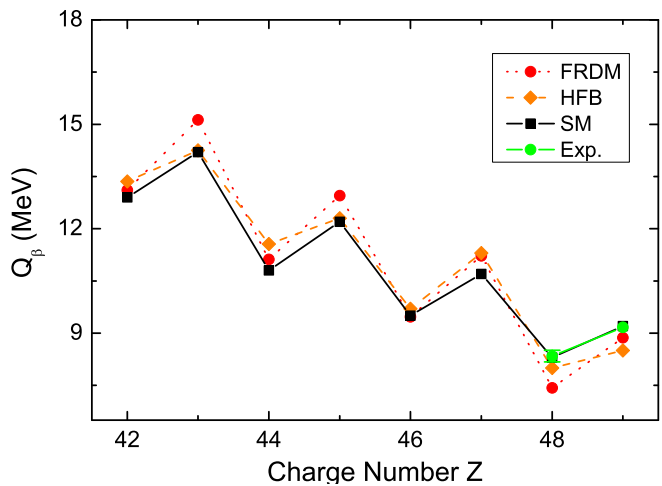}
  \caption{(Color online) Comparison of $Q_{\beta}$ values of the
    $N=82$ isotones between theoretical results from the FRDM
    \cite{Moellerff}, HFB \cite{Engel}, the present shell model
    approaches and experimental data \cite{Nndc}.}
  \label{fig:Q82}
\end{figure}

The present interaction and model space, based on a $^{88}$Sr core,
are not the same as used in Ref. \cite{Cuenca07} to calculate $Q_\beta$ values
and Gamow-Teller strength functions. However, we stress that the present
shell model calculation gives very similar results to the ones of Ref.
\cite{Cuenca07}. In particular we reproduce the experimentally
available $Q_\beta$ values very well, as is shown in Fig. \ref{fig:Q82}.
This figure also shows that the agreement of the $Q_\beta$ values obtained
in other models is usually not as good as by the shell model results. We will
in the following use the shell model $Q_\beta$ values for the calculation of
the half-lives.

The $1/2^-$ isomer in $^{131}$In corresponds approximately to a
$^{132}$Sn configuration with a hole in the $1p_{1/2}$ orbital.  Our
calculation reproduces the energy of the isomer at 0.302 MeV.  (This
quantity was one of the experimental ingredients to which the
interaction has been adjusted.)

\begin{figure}[htb]
    \includegraphics[width=\linewidth]{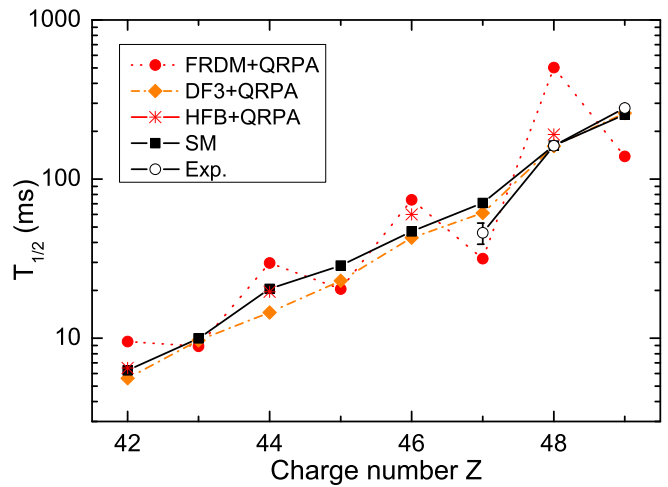}
    \caption{(Color online) Comparison of half-lives of the $N=82$
      isotones as calculated in the FRDM, HFB, DF3+QRPA
      \cite{Borzov06} and the present shell model approaches with
      data.}
    \label{fig:tau82}
\end{figure}

\begin{table}[htb]
  \caption{Comparison of the present shell model
    half-lives and the ones of reference~\cite{Cuenca07} with 
    experiment~\cite{Dil03,Fog04}. All
    half-lives are in ms.}
  \label{tab:lives82}
  \renewcommand{\arraystretch}{1.1}
  \begin{ruledtabular}
    \begin{tabular}{cccc}
      Nucleus & \multicolumn{3}{c}{Half-Life (ms)} \\ \cline{2-4} &
Expt.  & \textrm{present} & \textrm{shell model
(ref~\cite{Cuenca07})} \\ \hline
$^{131}$In  & $280 \pm 30$ & 247.53   &     260  \\
$^{130}$Cd  & $162 \pm 7$  & 164.29   &     162  \\
$^{129}$Ag  & $46^{+5}_{-9}$ & 69.81   &     70     \\
$^{128}$Pd  &              & 47.25    &      46  \\
$^{127}$Rh  &              & 27.98 &      27.65  \\
$^{126}$Ru  &              & 20.33 &    19.76  \\
$^{125}$Tc  &              & 9.52  &     9.44  \\
$^{124}$Mo  &              & 6.21  &     6.13   \\
    \end{tabular}
  \end{ruledtabular}
\end{table}

\begin{figure}[htb]
    \includegraphics[width=\linewidth]{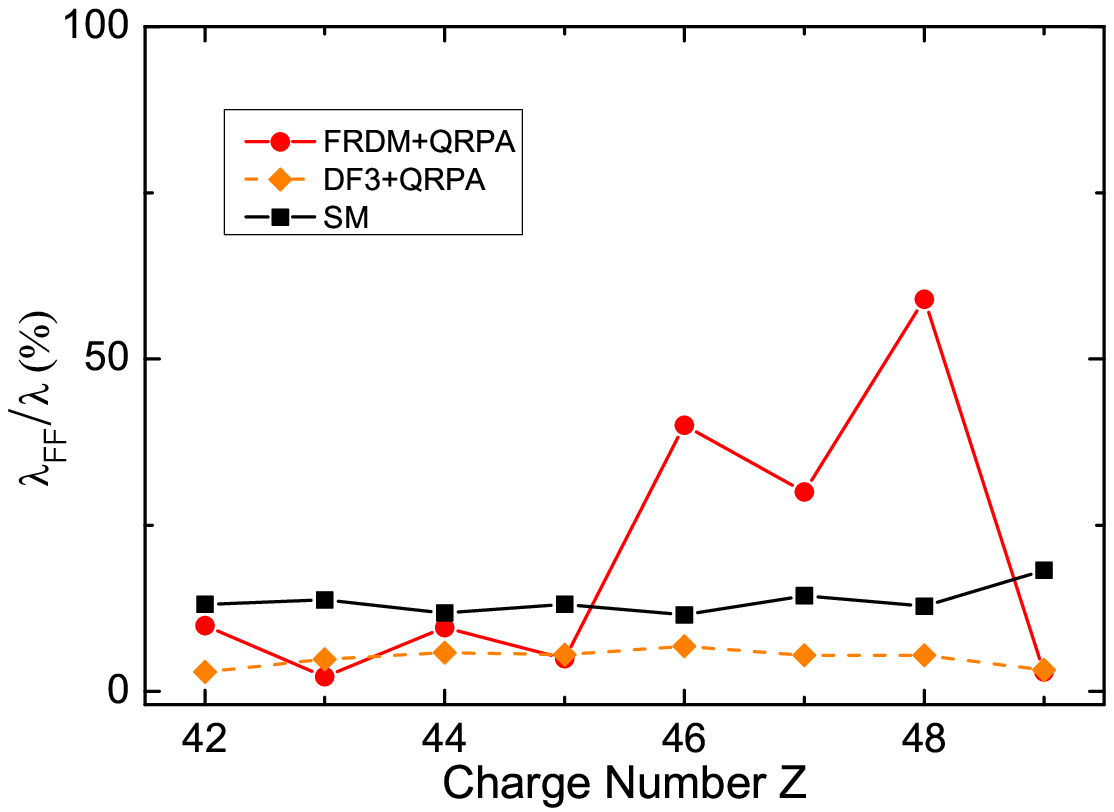}
    \caption{(Color online) Percentage of the contributions from
      first-forbidden transitions to the half-lives of the N$=82$
      isotones from FRDM+QRPA \cite{Moellerff}, DF3+QRPA
      \cite{Borzov,Borzov06} and the present shell model data.}
    \label{fig:rat82}
\end{figure}

\begin{figure*}
  \includegraphics[width=\linewidth]{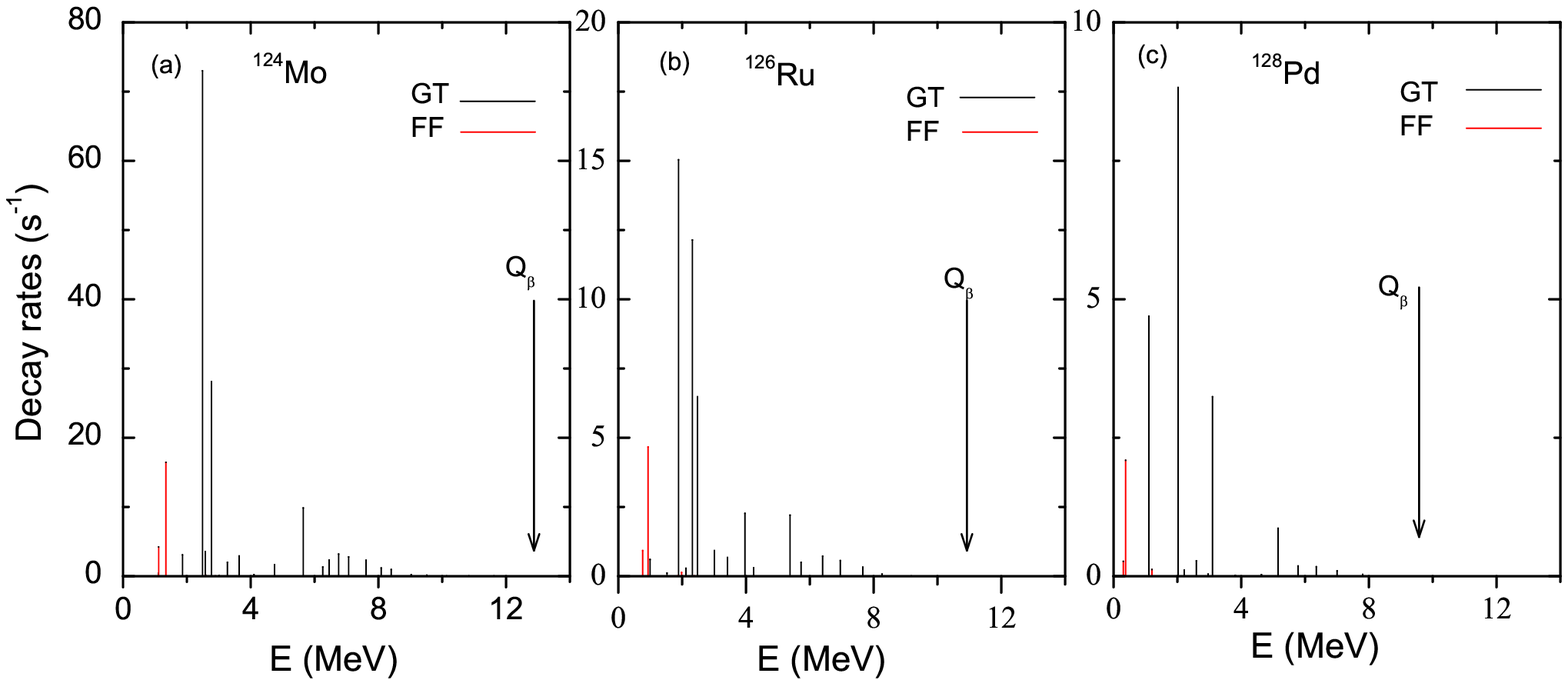}
  \caption{(Color online) Partial decay rates including GT and FF
    transitions for the $N=82$ isotones $^{124}$Mo (a), $^{126}$Ru (b)
    and $^{128}$Pd (c).\label{fig:rats-82}}
\end{figure*}

The calculated half-lives for the $N=82$ waiting point nuclei are
summarized in Table \ref{tab:lives82} and are compared to data and to
previous theoretical estimates in Fig. \ref{fig:tau82}.  Compared to
experiment, the half-life of $^{131}$In is well reproduced, while the
one for $^{129}$Ag is somewhat too long. This shortcoming had already
been observed in the previous shell model calculations. In fact, the
present shell model results, including contributions from
first-forbidden transitions, agree very well with the shell model
results of Ref. \cite{Cuenca07}. This, however, does not mean that
first-forbidden transitions are negligible. As is shown in Fig.
\ref{fig:rat82} first-forbidden transitions contribute about $13 \%$
to the half-life. However, this value is nearly the same for all
$N=82$ waiting point nuclei explaining the similarity between the
present shell model results to those of Ref. \cite{Cuenca07}. Only for
$^{131}$In, first-forbidden transitions contribute somewhat more,
resulting in a slightly smaller half-life than in the shell model
study based solely on Gamow-Teller transitions. We note that our
prediction of an $18\%$ contribution stemming from first-forbidden
transitions to the decay of the $^{131}$In ground state is in
agreement with the experimental limit of $\le 20\%$. We further add
that we calculate a Gamow-Teller contribution to the half-life of the
$1/2^-$ isomer in $^{131}$In which is less than $1\%$, confirming our
assumption to fix the quenching of the first-forbidden transition to
this decay.  Fig. \ref{fig:rats-82} shows the partial decay rates to
different final states for the nuclei $^{124}$Mo, $^{126}$Ru and
$^{128}$Pd. We note that Gamow-Teller transitions are larger than
first-forbidden transitions, which, however, proceed to levels at
lower excitation energies which enhances them by phase space.

\begin{figure}[htb]
    \includegraphics[width=\linewidth]{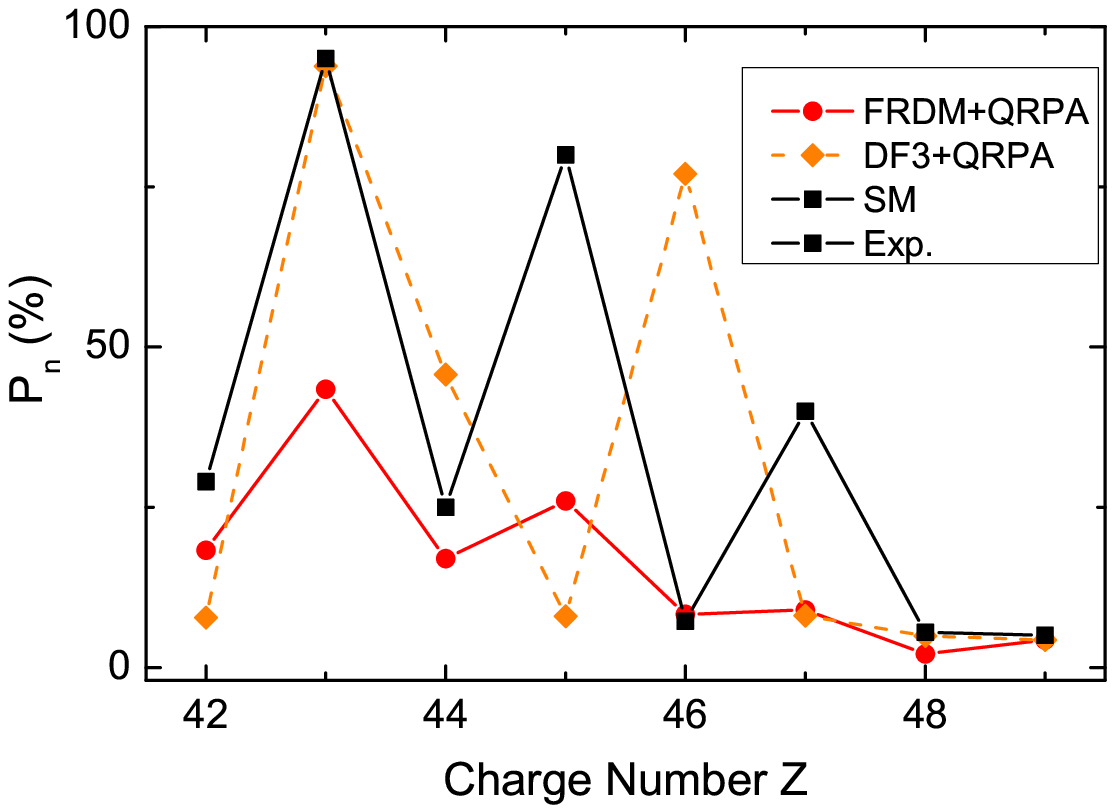}
    \caption{(Color online) $\beta$-delayed neutron emission
      probability for selected $N=82$ r-process nuclei from FRDM+QRPA
      \cite{Moellerff}, DF3+QRPA \cite{Borzov,Borzov06}, the present
      shell model and experiment \cite{Nndc}.}
    \label{fig:Pn82}
\end{figure}

The $\beta$-delayed neutron emission probabilities, i.e. the
probabilities that the decay leads to states in the daughter nucleus
above the neutron separation threshold and hence is followed by the
emission of a neutron, is obviously sensitive to a good description of
both the neutron separation energies and the $\beta$ strength
functions in the $Q_\beta$ window. As has been stressed in
Refs. \cite{Martinez99,Cuenca07} the improved description of
correlations in shell model calculations gives a more realistic
account of the fragmentation of the strength function than is obtained
in QRPA studies. Fig. \ref{fig:Pn82} compares the present shell model
probabilities to those obtained in Ref. \cite{Cuenca07}.  We find that
the inclusion of first-forbidden transitions leads only to minor
changes. A detailed comparison of the shell model results
\cite{Cuenca07} to those obtained in other theoretical approaches is
given in \cite{Martinez99,Cuenca07}.

\section{Half-lives of the $N=126$ waiting point nuclei}

\begin{figure}
    \includegraphics[width=\linewidth]{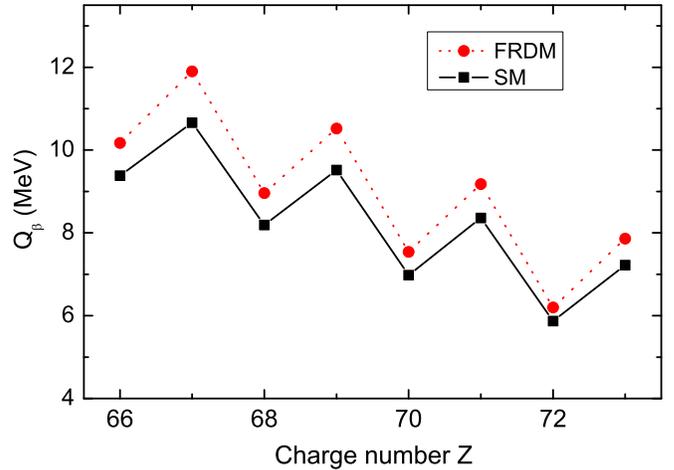}
    \caption{(Color online) Comparison of $Q_{\beta}$ values of the
      $N=126$ isotones as calculated in the FRDM \cite{Moellerff} and
      the present shell model approaches.}
    \label{fig:Q126}
\end{figure}

Fig. \ref{fig:Q126} compares the calculated $Q_\beta$ values of the
N=126 isotones with other theoretical models.  While the general trend
of the $Q_\beta$ is quite similar than obtained in the FRDM model, the
shell model values are slightly smaller than those from the FRDM
model. As, however, no experimental data exist for these very
neutron-rich $N=126$ nuclei, it is not possible to decide which
$Q_\beta$ are more realistic. In the following we will use the shell
model values to calculate the half-lives and $\beta$-delayed neutron
emission probabilities for the $N=126$ r-process waiting point nuclei.

\begin{table}[bth]
  \caption{Comparison of the present shell model half-lives and the
    one of reference~\cite{Suzuki12}. All 
    half-lives are in ms.}
  \label{tab:lives126}
  \renewcommand{\arraystretch}{1.1}
  \begin{ruledtabular}
    \begin{tabular}{ccccc}
      Nucleus & \multicolumn{3}{c}{Half-Life (ms)} \\ \cline{2-4}
      & \textrm{present} & \textrm{SM (ref~\cite{Suzuki12})}    \\ \hline
      $^{199}$Ta  & 286.17   & 278.88  \\
      $^{198}$Hf  & 193.28   & 129.65  \\
      $^{197}$Lu  & 107.85   & 84.81   \\
      $^{196}$Yb  & 68.98    & 44.18    \\
      $^{195}$Tm  & 36.03    & 29.49   \\
      $^{194}$Er  & 24.58    & 18.11   \\
      $^{193}$Ho  & 13.58    & 10.94   \\
      $^{192}$Dy  & 10.10    & 7.75   \\
    \end{tabular}
  \end{ruledtabular}
\end{table}

\begin{figure}[htb]
    \includegraphics[width=\linewidth]{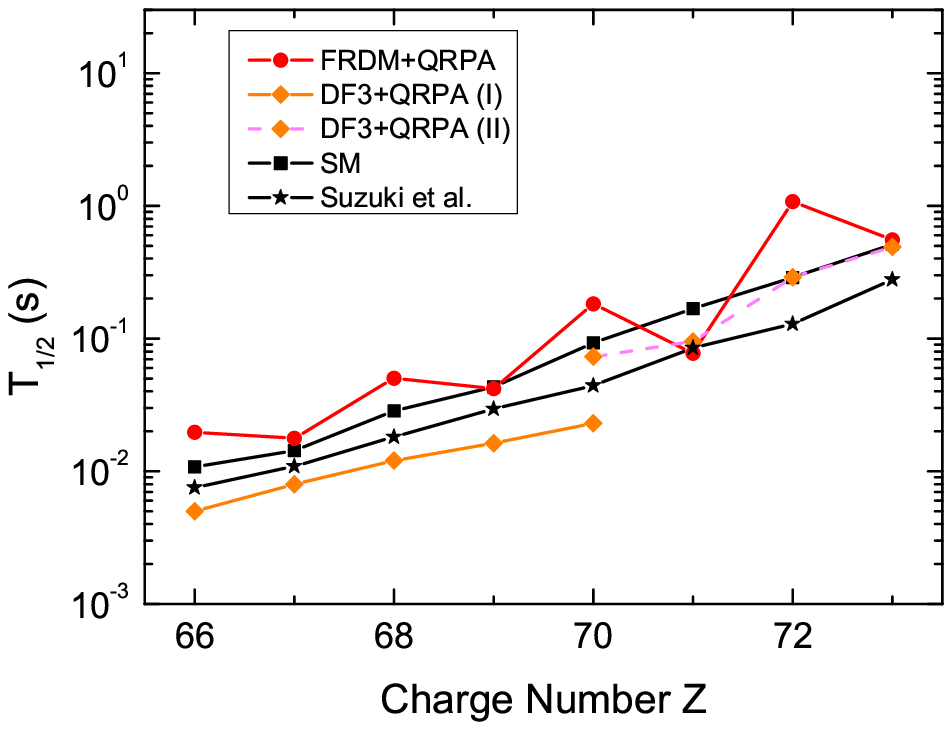}
    \caption{(Color online) Comparison of half-lives of the $N=126$
      isotones as calculated in the FRDM+QRPA, DF3+QRPA(I)
      \cite{Borzov06}, DF3+QRPA(II) \cite{Borzov11} and the present
      shell model approaches \cite{Suzuki12}.}
    \label{fig:lives126}
\end{figure}

The shell model half-lives are listed in Table \ref{tab:lives126} and
are compared to other theoretical predictions in
Fig. \ref{fig:lives126}.  Although recently researchers at GSI have
been successful to measure half-lives of nuclei close to $N=126$ with
charge numbers below lead \cite{Nieto}, supplying important
constraints about the half-life trend towards the r-process nuclei,
experimental data for the $N=126$ r-process nuclei do yet not exist.
Hence our results can only be compared to other theoretical
predictions. We note that the present half-lives for $Z>70$ are
faster, by about a factor of two, than those obtained by Borzov within
an QRPA approach on top of the density functional DF3, showing,
however, a similar dependence with charge number \cite{Borzov}.
Adopting a different parameterization, Borzov has also calculated
half-lives for $N=126$ isotones with $Z <70$ which are slightly faster
than the shell model values \cite{cqrpa}.  The shell model half-lives
are noticeably faster than those predicted previously by global
models, e.g. by the QRPA calculation on top of the
microscopic-macroscopic FRDM or ETFSI approaches.

Recently Suzuki \emph{et al.} \cite{Suzuki12} have presented the 
first shell model half-lives for $N=126$ r-process nuclei, including both 
GT and first-forbidden contributions. However, our present model 
space including the ($0g_{7/2}$, $1d_{5/2,3/2}$, $0h_{11/2}$, $2s_{1/2}$) 
proton orbits is noticeably larger than the one used in Ref. \cite{Suzuki12} 
(the $1d_{3/2}$, $0h_{11/2}$, $2s_{1/2}$ proton orbits). 
Relatedly the two shell model calculations differ in the
residual interaction and additionally in the adopted quenching scheme for 
first-forbidden transitions. 
Nevertheless, as is shown in Table 
\ref{tab:lives126} and Fig. \ref{fig:lives126}, 
both shell model calculations predict very similar half-lives for
the $N=126$ nuclei.
Both studies do not predict the strong odd-even staggering in
the half-lives as observed in the QRPA results on top of the FRDM model.

\begin{figure}[htb]
    \includegraphics[width=\linewidth]{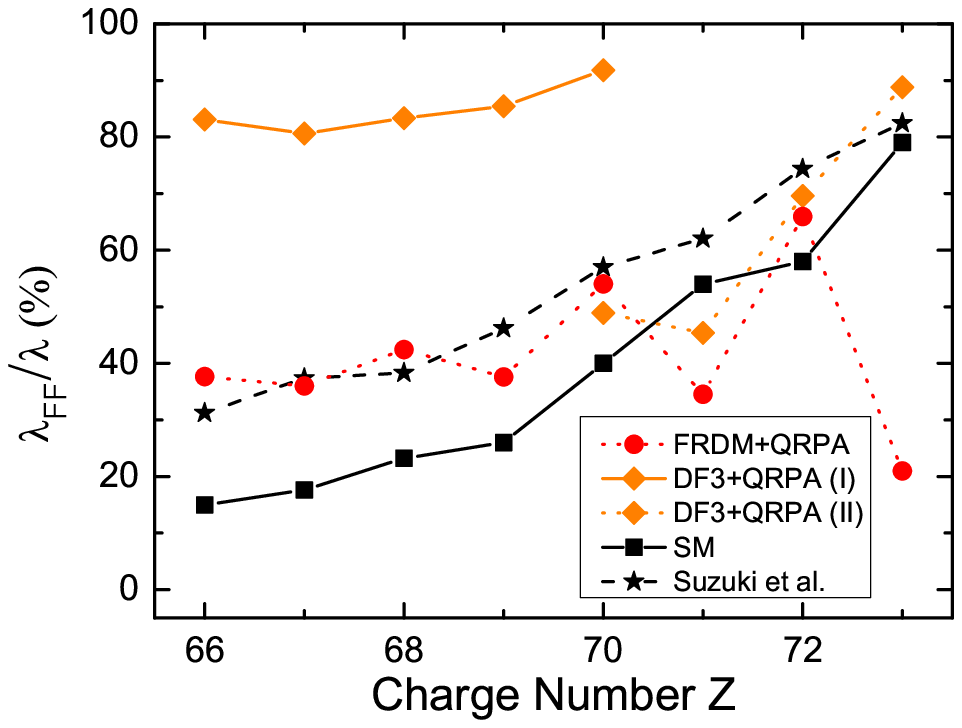}
    \caption{(Color online) Percentage of the contributions from
      first-forbidden transitions to the half-lives of the N$=126$
      isotones are compared with results from DF3+QRPA(I)
      \cite{Borzov06}, DF3+QRPA(II) \cite{Borzov11} and shell model
      approaches \cite{Suzuki12}.}
    \label{fig:contri126}
\end{figure}

\begin{figure*}[htb]
    \includegraphics[width=\linewidth]{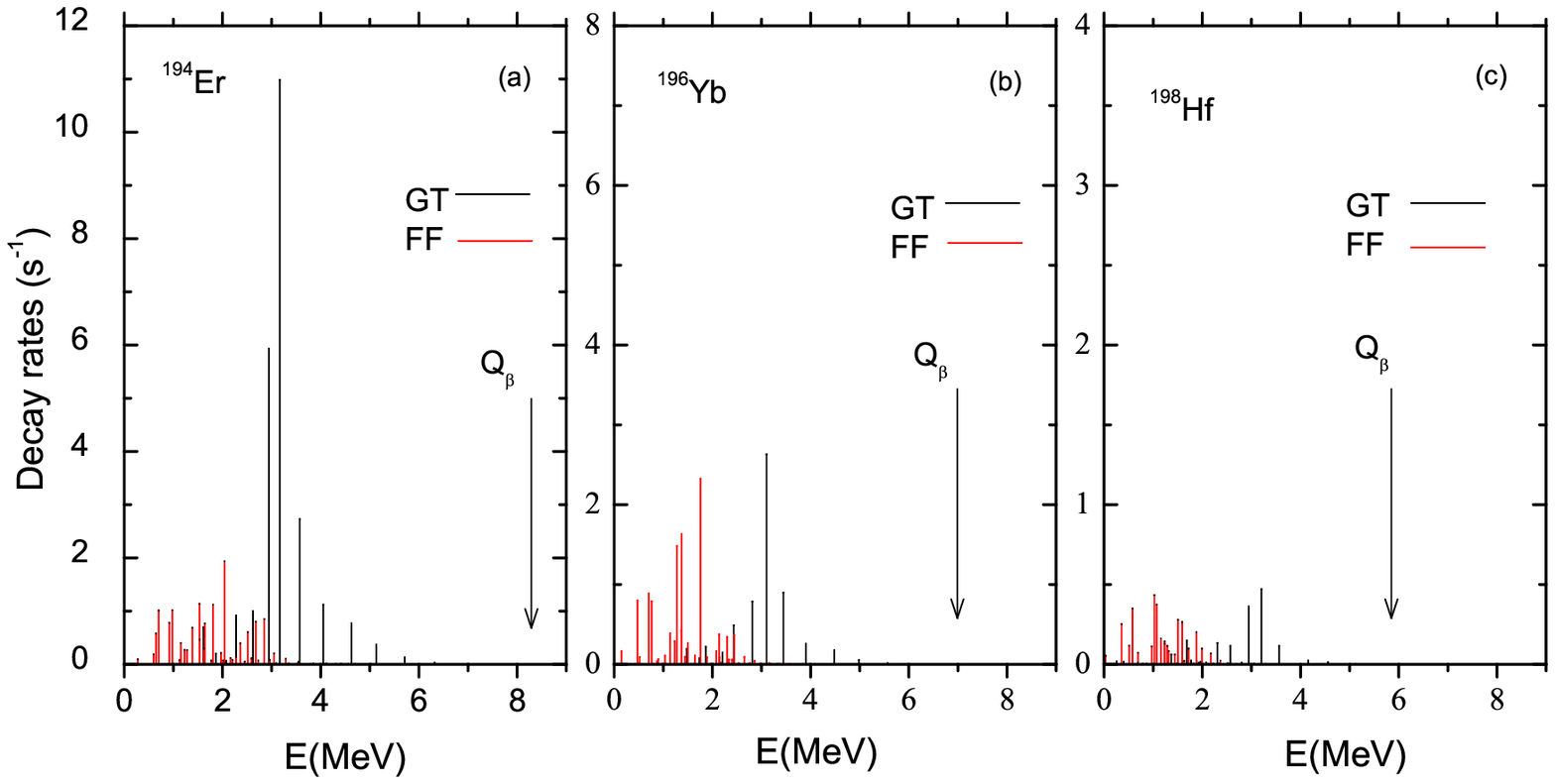}
    \caption{(Color online) Partial decay rates including Gamow-Teller
      and FF transitions for the $N=126$ isotones $^{194}$Er
      (a), $^{196}$Yb (b), and $^{198}$Hf (c). \label{fig:rates-dis-126}} 
\end{figure*}

As already noted in Ref. \cite{Borzov06} based on the density
functional calculations, first-forbidden transitions are expected to
contribute significantly to the half-lives of the $N=126$ r-process
nuclei. This finding is supported by our shell model calculations
(Fig. \ref{fig:contri126}).  One observes an increasing contribution
from the first-forbidden transitions with increasing proton number.
In fact, for nuclei with proton number $Z \ge 70$, contributions from
first-forbidden transitions to the half-life dominate over
Gamow-Teller transitions.  This behavior can be understood by
inspecting the partial decay rates arising from Gamow-Teller and
first-forbidden transitions which are shown in
Fig. \ref{fig:rates-dis-126} for selected nuclei.  We note that
Gamow-Teller transitions are related to the change of a neutron in the
$0h_{9/2}$ orbit to a proton in the $0h_{11/2}$ orbit which, however,
is fragmented over several states in the daughter nucleus due to
correlations.  Nevertheless, for the nuclei studied here these final
proton states reside at moderately high excitation energies around 3
MeV, while first-forbidden transitions connect to excited states at
lower excitation energies.  With increasing proton number, more
protons occupy the final $0h_{11/2}$ orbit and the GT transitions get
gradually Pauli blocked.  This explains why the GT strength gets
strongly reduced with increasing proton number. Actually also
first-forbidden transitions get blocked with increasing proton number,
which is, however, a significantly milder effect as for the GT
transitions.  We note that for all nuclei studied here first-forbidden
transitions are mainly mediated by rank 0 and 1 operators (the latter
contributes about $70\%$ to the forbidden strength), while the
contribution arising from rank 2 operators are very small. Due to its
larger sensitivity to phase space ($\sim Q^7$) the relative
contribution of rank 2 transitions increases slightly with decreasing
charge number.  For the $N=126$ nuclei studied here, the depopulation
of the $0h_{11/2}$ proton orbital with decreasing charge number, which
increases both the Gamow-Teller and first-forbidden transitions,
dominates the trend observed in our half-lives.  Changes in phase
space and pairing have lesser effects on our half-life systematics.

\begin{figure}[htb]
    \includegraphics[width=\linewidth]{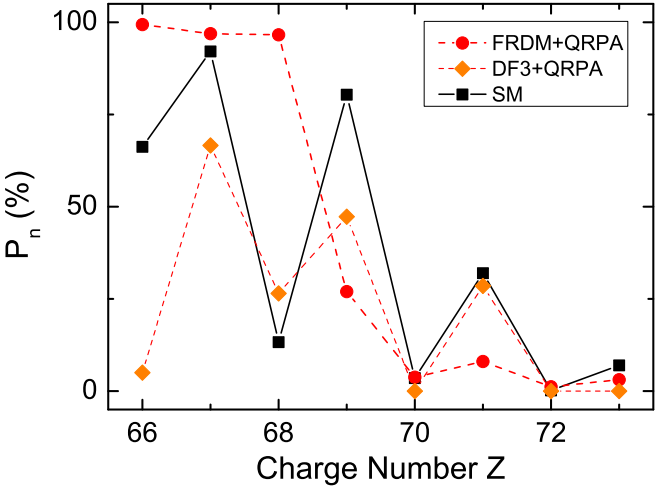}
    \caption{(Color online) Neutron emission probability of $N=126$.}
    \label{fig:prob126}
\end{figure}

Gamow-Teller transitions proceed to final states mainly above the
neutron threshold and hence are accompanied by neutron emission, while
the final states populated by first-forbidden transition predominantly
reside below the neutron threshold. Hence we expect from the
$Z$-dependence of the GT and first-forbidden transitions that the
$\beta$-delayed neutron emission probability decreases with increasing
proton number. This is indeed confirmed by Fig. \ref{fig:prob126}. The
striking odd-even staggering is related to pairing which reduces the
neutron threshold energies in odd-odd daughter nuclei relatively to
odd-$A$ nuclei, but does basically not affect the strength
distributions, as discussed in \cite{Langanke00}. The QRPA/FRDM
neutron emission probabilities show a rather abrupt increase at $Z=68$
which is likely due to the fact that QRPA calculations show
significantly less fragmentation of the strength than shell model
studies and that, for $Z<69$, the few dominant transitions reside
above the neutron threshold.

\section{SUMMARY AND CONCLUSIONS}

We have calculated the half-lives and $\beta$-delayed neutron emission
probabilities of the r-process waiting point nuclei with magic neutron
numbers $N=50, 82$, and 126 within the framework of the large-scale
shell-model.  The calculations include contributions both from allowed
Gamow-Teller and first-forbidden transitions. We find good agreement
with the existing experimental data: i.e. the half-lives for the
$N=50$ nuclei with charge numbers $Z=28$-32 and for the $N=82$ nuclei
$^{129}$Ag and $^{130}$Cd.  In our calculations first-forbidden
transitions significantly reduce the half-lives of the $N=126$ waiting
point nuclei, while they have a smaller effect on the half-lives of
the $N=50$ and 82 r-process nuclei.

\begin{acknowledgments}
  This work was supported by the Helmholtz International Center for
  FAIR within the framework of the LOEWE program launched by the state
  of Hesse, the Helmholtz Association through the Nuclear Astrophysics
  Virtual Institute (VH-VI-417), the ExtreMe Matter Institute EMMI in
  the framework of the Helmholtz Alliance HA216/EMMI, the Deutsche
  Forschungsgemeinschaft through contract SFB 634, the National
  Natural Science Foundation of China (11165006, 11105079) and 
  the IN2P3-GSI collaboration agreement (10-63). Discussions with
  Fr{\'e}d{\'e}ric Nowacki are greatfully acknowledged.
\end{acknowledgments}

\end{document}